\documentstyle[preprint,eqsecnum,aps]{revtex}

\begin{document}
\draft
\title{Density Perturbations of Thermal Origin During Inflation}
\author{Wolung Lee and Li-Zhi Fang}

\address{Department of Physics, University of Arizona, Tucson, AZ 85721}
\date{\today}
\maketitle
\begin{abstract}

We study thermally induced density perturbations during inflation.
This scenario is characterized by two thermodynamical conditions: (1)
The primordial perturbations originate in the epoch when the
inflationary universe contains a thermalized heat bath. (2) The
perturbations of the inflationary scalar field are given by the
fluctuation-dissipation relation. We show that the  spectrum of the
primordial perturbations is of power law, but tilted, and there is a 
relation between the amplitude and the index of the power spectrum.
Aside from the mass scale of the inflation, the amplitude-index 
relation does not depend on other parameters like $g$-factor. These 
results are found to be well consistent with observations of the 
temperature fluctuations of cosmic microwave background if the mass 
scale of the inflation is about $10^{15}$ GeV. Instead of the purely
adiabatic case, the consequent density perturbation is an admixture of 
adiabatic and isocurvature one.  Therefore, the detection of 
super-Hubble suppression of the spectrum would be effective for further 
discrimination between the thermally originated models and others.

\end{abstract}


\narrowtext

\section{Introduction}
\label{sec:level1}

Recently, the possibility of thermally induced initial density perturbations 
in an inflationary cosmology was examined \cite{BF}. It was found that
for a wide parameter range within the standard inflation model, the thermal 
fluctuations of the scalar field can play an important and even dominant role 
in producing the initial perturbations.

The ``standard" scenario of inflation \cite{KT} assumes that during the 
inflation the cosmic expansion was dominated by the vacuum energy of a scalar
field $\phi$, and the thermalized component has never existed, or if a thermal
component like radiation existed at the onset of inflation, it was blown
off by the exponential expansion. The thermal component was not established or 
re-established until heating or reheating caused by the decay and 
thermalization of the $\phi$ field into relativistic particles. The thermal 
inflation scenario also assumes that the inflation is dominated by a scalar 
field $\phi$.  Unlike the standard model, the thermal scenario requires that a 
thermalized component be established or re-established before the end of the 
inflation. Therefore, the thermal model can be considered similar to the 
``standard" model, but with a heating or reheating stage of the $\phi$-decayed 
particles while the inflation was still in progress.

It has been realized that if the interaction of the inflation field with
light  particles is weak, namely, of the order of gravitational strength or
slightly stronger, the thermal component cannot maintain itself during the
inflation. In contrast, if the interaction is strong enough to lead to 
thermalization, it is possible to maintain a heat bath in the epoch of 
inflation \cite{AB}.  Therefore, the thermal scenario probably can be ``
installed" onto various versions of inflation by adding a stronger interaction 
between the $\phi$ field and the heat bath.

In this paper, we study the properties of density fluctuations of thermal 
origin.  These density perturbations are characterized by
two thermodynamical conditions: (1) the perturbations  originate in the 
epoch when the inflationary universe contains a thermalized heat bath;
(2) the perturbations of the inflationary scalar field are given by the 
fluctuation-dissipation relation. Therefore, one can expect that, like many
problems of thermal physics, some features of the perturbations should be 
mainly determined by thermodynamical requirements, but less dependent
on the details of the model, such as the inflationary potential. Obviously, 
these features would be useful in testing the thermal scenario of perturbation 
production, and discriminating it from other models.

Up to now, the only possible way of directly testing models of the origin
of cosmic inhomogeneity is to compare the predicted spectrum of the
perturbations with the observed temperature fluctuations of the cosmic
microwave background (CMB). Therefore, we will focus on the observable
features of the perturbations, including the power law index, the magnitude
of the spectrum, and the behavior of the perturbations on scales larger
than the Hubble radius.

In Sec. II we discuss the exact and approximate solutions of the evolution of 
the radiation component in the thermal inflation scenario.  Sec. III
gives the calculation of the features of the density perturbations produced
during the thermal inflation. Finally, the observational tests and possible
discrimination between this model and others are discussed in Sec. IV.

\section{Heating or Reheating During Inflation}

\subsection{Thermal inflation scenario}

The dynamics of a universe consisting of an inflationary scalar field
$\phi$ and a thermalized radiation component can be described by
the following three equations\cite{BF}. First, the equation of
motion of the $\phi$ field is
\begin{equation}
\ddot{\phi} + 3H \dot{\phi} + \Gamma_{\phi}\dot{\phi} 
+ V'(\phi)=0 \ .
\end{equation}
where $V(\phi)$ is the effective potential. Second, the
equation of the radiation component (heat bath) is given by the
first law of thermodynamics as
\begin{equation}
\dot \rho_{r} + 4H\rho_{r} = \Gamma_{\phi}\dot{\phi}^{2},
\end{equation}
where $\rho_{r}$ is the energy density of the thermal component.
The temperature of the thermal bath can be calculated by
$\rho_{r}= (\pi^{2}/30) g_{\rm eff}T^{4}$, $g_{\rm eff}$ being the effective
number of degrees of freedom at temperature $T$. The friction terms of
$\Gamma_{\phi}$ in Eqs. (2.1) and (2.2) phenomenologically describe the
interaction between $\phi$ field and the heat bath (radiation component)
\cite{AL}.
The third equation is the Hubble ``constant" $H ={\dot R}/R$, {\it i.e,}
\begin{equation}
 H^{2} =
\frac{8\pi}{3}\frac{1}{m_{\rm Pl}^2}[\rho_{r} + 
\frac{1}{2}\dot\phi^{2} + V(\phi)],
\end{equation}
where $m_{\rm Pl}\equiv 1/\sqrt{G}$ is the Planck mass.

In the standard scenario of inflation, the radiation component is assumed
to be nonexistent, or to be blown away by the exponential expansion, and the
universe is cool $(T \sim H)$ during the inflation era.  To establish or 
re-establish a radiation component, a heating or reheating epoch following the 
inflation is needed.  However, if the friction coefficient $\Gamma_{\phi}$ is 
comparable to or larger than $H$, and there is strong coupling between the 
decay products of the scalar field, then the system of the decaying products
will quickly reach thermal equilibrium. In this case, a non-zero and
thermal component $\rho_r$ can be maintained during the inflation.

One may be concerned that the friction term
$\Gamma_{\phi}\dot{\phi}$ is not appropriate for describing the energy 
transfer via $\phi$ field particle production under far-from-equilibrium 
conditions. Indeed, it has been shown that in some models of reheating, the 
first stage of the $\phi$ field decaying into other bosons is due to a 
parametric resonance \cite{KLS}.  However, in the last stage while the 
thermalization is established the friction mechanism prevails. Therefore,
the
description of $\Gamma_{\phi}$ is reasonable if there is a thermalized heat 
bath \cite{AB}.

Briefly, thermal inflation scenario can be characterized by the following
properties:\\
1.) The relaxing time scale of the heat bath is shorter than the expansion
\begin{equation}
\Gamma_{\phi} > H,
\end{equation}
2.) The energy of the universe is dominated by $\phi$ field potential
\begin{equation}
V(\phi) \gg \dot{\phi}^2/2, \mbox{\ \ \ \ \ \ }  V(\phi) \gg \rho_r.
\end{equation}
3.) The temperature of the heat bath is larger than $H$,
\begin{equation}
T > H.
\end{equation}
4.) As in the standard inflation scenario, Eq. (2.1) should have a slow-roll 
solution.  That is, the potential of the inflaton $\phi$ field is ``flat",
\begin{equation}
|V'(\phi)| \ll (3H+\Gamma_{\phi})V^{1/2}(\phi),
\end{equation}
and
\begin{equation}
|V^{''}(\phi)| \ll (3H+\Gamma_{\phi})^2.
\end{equation}
As such, the slow-roll solution of Eq. (2.1) is given by
\begin{equation}
\dot{\phi} \simeq - \frac{V'(\phi)}{3H + \Gamma_{\phi}}.
\end{equation}

It has been shown that conditions (2.4) - (2.8) can  hold
simultaneously if the coupling constant $\Gamma_{\phi}$ and the potential
$V(\phi)$ of inflation field satisfy \cite{BF}
\begin{equation}
\left(\frac{M}{m_{\rm Pl}}\right)^5\frac{M}{W}< \Gamma_{\phi} < \left(\frac{M}
{m_{\rm Pl}}\right)\frac{M}{W},
\end{equation}
\begin{equation}
V^{3/2}(\phi)m_{\rm Pl}^{-3} \ll V'(\phi) \ll m_{\rm Pl}^{-1}V(\phi).
\end{equation}
where $M$ is the mass scale of the inflation, and
$W = \dot{\phi}^2/2V(\phi)$ is the ratio of the kinetic and potential
energy of the $\phi$ field. In the case of $(M/m_{\rm Pl})^2 \ll 1$, there
is a large region in the parameter space of the potential fulfilling the
conditions Eqs. (2.10) and (2.11).

Exactly speaking, we should use a finite temperature effective potential 
$V(\phi,T)$ in the dynamics of thermal inflation. However, for many popular 
potentials, the correction due to finite temperature is negligible. 
As an example, we consider a $\phi^4$ potential given by
\begin{equation}
V(\phi)=\lambda(\phi^{2}-\sigma^{2})^{2}.
\end{equation}
The slow-roll solution gives
\begin{equation}
H^2 \simeq H_i^2 \equiv \frac{8\pi}{3}\frac{V(0)}{m_{\rm Pl}^2},
\end{equation} 
where $V(0)\equiv M^{4}=\lambda \sigma^{4}$, and the subscript $i$ denotes
the time of the beginning of the inflation. The condition (2.8) can
certainly be satisfied if $\lambda \leq (M/m_{\rm Pl})^4$. On the other
hand, the leading temperature correction of the potential (2.12) is
$\lambda T^2 \phi^2$. Therefore,
$\lambda T^2 \leq M^6/m_{\rm Pl}^4 \sim (M/m_{\rm Pl})^2H^2 \ll H^2$, 
{\it i.e,} the influence of the finite temperature effective potential is 
negligible when $\phi < m_{\rm Pl}$.

\subsection{Evolution of Radiation Component}

The evolution of the field in the thermal inflation scenario is about the
same as in the standard model.  However, the radiation component shows very
different behavior in these two scenarios. In the standard model, the
radiation component did not exist before reheating, while in the
thermal scenario, it appeared quite early. To illuminate this point, we
first plot a typical numerical solution of Eqs. (2.1) - (2.3) in Fig. 1,
where the potential $V(\phi)$ is given by (2.12) and the relevant parameters
are taken to be $M = 10^{15}$ GeV, $\lambda = 4 \cdot 10^{-17}$,
$\Gamma_{\phi} = 15 H_i$ and $g_{\rm eff}$ = 100. It should be pointed out that
$g_{\rm eff}$-factor generally is a function of $T$, and the unknown function
$g_{\rm eff}(T)$ will lead to uncertainty in the solutions. Fortunately, 
the variation of $g_{\rm eff}$-factor has only a slight effect on the problems 
we are going to study.

The solution shown in Fig. 1 is easy to understand. As expected,
the inflation is dominated by the vacuum energy of the $\phi$ field, and
$H$ remains almost constant during the inflation. The evolution of the
radiation contains three phases. Phase 1 covers the period during which the 
radiation temperature $T$ drops drastically due to the inflationary expansion.
This phase is the same as in cool inflation models. When $t=t_b$, $T$ will 
rebound, and the evolution of radiation evolves into phase 2 which lasts from 
$t_b$ to $t_f$. In this phase, $H \sim H_i$, while the radiation temperature
increases due to the friction of the $\phi$ field. Therefore, the second phase 
is actually the epoch of inflation plus reheating. Finally, phase 3 begins at
$t = t_f$ when the energy of the radiation components is comparable to the
potential energy of $\phi$ field, and the inflationary process stops.
The third phase is, in fact, the standard radiation dominated universe, and
radiation temperature follows the standard evolution of adiabatic
decreases with the expansion of the universe.

Therefore, the only difference between the thermal and the standard
scenarios is that the ``reheating" starts at the time $t_b < t_f$ when the
inflation still prevails. In other words, the reheating stage is merged into 
the epoch of inflation. A specific reheating stage between the epochs of 
inflation and radiation dominated universe is not needed. The transition from 
inflation to a radiation dominated universe is now smooth, because a heat 
bath has already been established before the end of the inflationary expansion.

Moreover, the solution of phase 2 is independent of the initial
conditions of the radiation and the $\phi$ field. Whatever the initial
condition of the radiation, the solution during phase 2 is always
about the same, as shown in Fig. 1, because the radiation in phase 2 is
produced during inflation, and the initial radiation has almost been consumed
by the expansion. Therefore, this solution of phase 2 can be installed in both 
initially hot and cool models. It can be used as either a reheating or heating 
solution.

The features shown in Fig. 1 are typical of thermal inflation. One can 
describe these features by approximate solutions of Eqs. (2.1)-(2.3). During 
the slow-roll stage we have $\phi \ll \sigma$, it would be reasonable to 
neglect the $\phi^{3}$ term in $V'(\phi)$. Eq. (2.1) is then
\begin{equation}
\ddot{\phi}+(3H + \Gamma_{\phi})\dot{\phi}-4\lambda\sigma^2 \phi=0.
\end{equation}
The inflationary solution of Eq. (2.14) can immediately be found as
\begin{equation}
\phi =\phi_i e^{\alpha Ht},
\end{equation}
where $\phi_i$ is the initial value of $\phi$ field. $\phi_i$ can be
estimated by $V'(\phi_i) = H_i^3$, which gives
$\phi_i \sim (M/m_{\rm Pl})^2 \sigma$. The uncertainty of $\phi_i$ slightly
affects the results discussed below.

The coefficient $\alpha$ in Eq. (2.15) is given by
\begin{equation}
\alpha=\frac{1}{2}\left(\sqrt{{\gamma}^{2}+
\frac{6}{\pi}\lambda^{1/2}\left(\frac{m_{\rm Pl}}{M}\right)^2}-{\gamma}\right),
\end{equation}
where $\gamma=(\Gamma_{\phi}+3H)/H_i$. Since we have required
$\lambda \leq (M/m_{\rm Pl})^4 \ll (M/m_{\rm Pl})^2$, and $\gamma \gg 1$,
Eq. (2.16) gives $\alpha \simeq 3\lambda^{1/2}(m_{\rm Pl}/M)^2/2\pi\gamma \ll
1$.

 From Eq. (2.15), one can find the exact solution of Eq. (2.2) as
\begin{equation}
\rho_r  = C'  e^{-4Ht} + C  e^{2 \alpha Ht},
\end{equation}
where $C=\alpha^{2}H\Gamma_{\phi}\phi_i^{2}/2(\alpha+2)$. If the initial 
condition is taken to be $\rho_r(0) = a M^4$, where $a = (\pi^2/30)g_{\rm eff}$
, we have $C'= \rho_r(0) - C$. It is clear from Eq. (2.17) that when the first
term on the right-hand side of Eq. (2.17) is dominant, the evolution is in
phase 1; and phase 2 starts when the second term becomes dominant. The transfer
from phase 1 to phase 2 occurs at the rebound of the temperature of 
radiation. The transition time $t_b$ can be determined by 
$(d\rho/dt)_{t_b}=0$. We have
\begin{equation}
Ht_b \simeq \frac{1}{2(\alpha+2)}
\ln \left(\frac{4(\alpha+2)}{\alpha^3H}
\frac {aM^4}{\Gamma_{\phi}\phi^2_i} \right ).
\end{equation}
The temperature at the rebound is
\begin{equation}
T_b = a^{-1/4}
(1+\frac{\alpha}{2})^{1/4}C^{1/4}e^{\frac{1}{2} \alpha H t_b}.
\end{equation}

Therefore, the radiation temperature $T$ in phases 1 ($t<t_b$) and
2 ($t>t_b$) can be approximately expressed as
\begin{equation}
T(t)  = \left \{ \begin{array}{ll}
                 T_b e^{-H(t-t_b)} \ \ \ \ &        \mbox{if \ $t < t_b$} \\
                 T_b e^{\frac{1}{2} \alpha H (t-t_b)} & \mbox{if \ $t > t_b$}
                 \end{array}
        \right.
\end{equation}
Substituting Eqs. (2.15), (2.18), and (2.19) into the approximate solution
(2.20) of $t > t_b$, we have
\begin{equation}
\rho_r =aT^4= \frac{\Gamma_{\phi}}{4} H \alpha^2 {\phi_i}^2 e^{2\alpha H t} =
 \frac {\Gamma_{\phi}}{4H} \alpha^2 H^2 \phi^2 = 
  \frac{\Gamma_{\phi}}{4H} \dot{\phi}^2.
\end{equation}
Substituting (2.21) into Eq. (2.2), we have
\begin{equation}
\dot \rho_r \sim 0.
\end{equation}
Solution (2.22) implies that the evolution of the radiation is simply
determined by the balance between the depletion of the initial radiation
density due to inflationary expansion and the production of thermal 
components by the friction.

 From Eq. (2.21), we have
$T = (4a)^{-1/4}\alpha^{1/2}\Gamma_{\phi}^{1/4}H^{1/4}\phi^{1/2}$.
The second phase will end at the time $t_f$, when $\phi$
approaches the minimum of $V(\phi)$, {\it i.e,} $\phi \leq \sigma/\sqrt{2}$.
Accordingly, the temperature at the end of phase 2 should be 
\begin{equation}
T_f \leq  (4a)^{-1/4}\alpha^{1/2}
\Gamma_{\phi}^{1/4}H^{1/4}(\sigma^2/2)^{1/4}
= \epsilon M,
\end{equation}
where
\begin{equation}
\epsilon \sim \left[ \frac{\alpha 
\Gamma_{\phi}}{2a (\Gamma_{\phi} + 3H)} \right]^{1/4}.
\end{equation}
Eq. (2.23) shows that the temperature of radiation will rise to of order $M$
when the inflation ends. Accordingly, both the heating (or reheating) and the 
inflation are underway in phase 2, and no post-inflation reheating is needed.

Using the same parameters as the numerical solution of Fig. 1, the
approximate solutions (2.20) and (2.23) are plotted. It shows that
this approximation is in excellent agreement with the numerical one.

\subsection{Duration of the thermal inflation}

Generally, the rebound temperature $T_b$ can be either greater or less
than $H$. Figure 2 plots a typical solution in the case of $T_b <H$.
This solution shows that the temperature $T$ is lower than $H$ in the
period of $t_b < t < t_e$, and it is higher than $H$, {\it i.e,} $T > H$ when
$t > t_e$. From Fig. 2, the evolution of $T$ at $T_b<H$ shows about the
same behavior as that in the period of $T_b \geq H$. However, we should not
consider the solution to be physical when $T<H$, because it is impossible to 
define a thermalized heat bath with temperature less than the Hawking 
temperature $H$. Nevertheless, this solution should be available when $t > t_e$
and $T > H$. 

As has been discussed in Sec. II.B, the evolution of radiation at $t > t_e$ 
is independent of whether the radiation existed at $t<t_e$. The behavior
of $T$ at the period $t > t_e$ is completely determined by the competition
between the diluting and producing radiation at $t>t_b$. Initial information
about the radiation has been washed out during the thermalization. For
instance, one can replace the solution at $t <t_e$ by a
standard (cool) inflation. Therefore, $t_e$ can be considered as the
time of the onset of thermal inflation, and the duration of the thermal 
inflation is $t_f - t_e$. The number of $e$-folds of the growth of the cosmic
factor $R$ during the thermal inflation is then
\begin{equation}
N \equiv \int_{t_e}^{t_f} Hdt \simeq \frac {2}{\alpha} \ln \frac {T_f}{H}.
\end{equation}

One can also formally calculate the number of $e$-folds of the growth in phase 
2 as 
\begin{equation}
N_2 \equiv \int_{t_b}^{t_f} H dt \simeq
\frac {2}{\alpha} \ln \frac {T_f}{T_b},
\end{equation}
and the number of $e$-folds of the total growth as 
\begin{equation}
N_t \equiv \int_{0}^{t_f} Hdt \simeq \frac{2}{\alpha}
\ln \left(\frac {\epsilon M}{T_b}\right) + Ht_b
=\frac{2}{\alpha} \ln \left[\left(\frac {2}{\alpha \gamma}\right)^{1/4}
    \frac {M}{\sqrt{\phi_i H}}\right].
\end{equation}
As expected, both $N_2$ and $N_t$ depend on the initial value of the field 
$\phi_i$, but $N$ does not. Therefore, $\phi_i$ will not lead to uncertainty
in our analysis if we only study the problems of thermal evolution at
$t_e < t < t_f$.   

\section{The primordial density perturbations}

\subsection{Fluctuations of the $\phi$ field}

The possibility for the fluctuations of thermal origin during inflation
is simply because there is a gap between the energy scales of the inflation 
$M$, and the Hawking temperature $H$. There is room for the temperature of 
the radiation component $T$ to be less than $M$ but larger than $H$. The 
condition $T < M$ ensures the condition Eq. (2.5), and then, the evolution of 
the $\phi$ field leads to an inflation. Condition $T > H$ is necessary in order
to have the perturbations be dominated by thermal fluctuations.

To calculate the thermal fluctuations, we rewrite Eq. (2.9) as \cite{BF}
\begin{equation}
\frac {d\phi}{dt} =  - \frac{1}{3H + \Gamma_{\phi}}
\frac{dF[\phi]}{d \phi},
\end{equation}
where $F[\phi] = V(\phi)$. This is, in fact, a rate equation of order
parameter $\phi$, describing a homogeneous system with the free energy 
$F[\phi]$ approaching to equilibrium. A rate equation like (3.1) cannot
correctly describe the approach to equilibrium during a phase transition 
without a noise term \cite{gold}. To ensure that the system approaches 
the global minimum, the order parameter dynamics is not pure relaxation 
but exhibits fluctuations, arising from the microscopic degrees of 
freedom. Namely, Eq. (3.1) should have an additional noise term 
$\eta$ such as
\begin{equation}
\frac {d\phi}{dt} =  - \frac{1}{3H + \Gamma_{\phi}}
\frac{\delta F}{\delta \phi} + \eta(t). 
\end{equation}
This is the Langevin equation for a system with one degree of freedom,
and the correlation functions of $\eta$ is determined by the
fluctuation-dissipation theorem \cite{FDT} as
\begin{equation}
\langle \eta(t)\eta(t')\rangle =
\frac {3H^3}{2\pi}\frac{T}{3H+\Gamma_{\phi}}.
\end{equation}
The fluctuations of $\phi$ field for $V''(\phi) \ll H^2$ are then
\begin{equation}
\langle(\delta\phi)^{2}\rangle = \frac{3}{4\pi}HT.
\end{equation}
Obviously, when $T>H$, the thermal fluctuations are larger than the quantum
ones. When $T<H$, the thermal component is negligible, and fluctuations
should be of quantum origin, {\it i.e,}
$\langle(\delta\phi)^{2}\rangle = \frac{3}{4\pi}H^2$.

Like all inflation models, the primordial perturbations undergo an
evolution of crossing outside, and later again crossing back inside the
Hubble radius. The density perturbations at re-entering the horizon are 
\begin{equation}
\left(\frac{\delta\rho}{\rho} \right)_h =
\frac{-\delta\phi V^{\prime}(\phi)}{\dot\phi^{2} + (4/3)\rho_{r}} \ .
\end{equation}
All quantities in the right-hand side of Eq. (3.5) are calculated at the
time when the relevant perturbations crossed out beyond the horizon at the
inflationary epoch. 

We are only interested in the primordial density perturbations produced
in the period of thermal inflation, {\it i.e,} $t>t_e$. From Eq. (3.5), we have
\begin{equation}
\left(\frac{\delta\rho}{\rho} \right )_h  \simeq
               \frac{9}{2}\left(\frac{5}{2\pi^3}\right)^{1/2}
   \left(\frac{\Gamma_{\phi}H^2}{g_{\rm eff}T^3}\right)^{1/2},  
\end{equation}
where the temperature $T$ should be in the range $T_f > T > H$.
 
\subsection{Power law index}

Since the inflation is immediately followed by the radiation dominated epoch, 
with no reheating in between, the comoving scale of a perturbation with 
crossing over (the Hubble radius) at time $t$ is given by
\begin{equation}
\frac {k}{H_0} = 2\pi \frac{H}{H_0} \frac{T_0}{T_f}e^{H(t-t_f)},
\end{equation}
where $T_0$ and $H_{0}$ are the present CMB temperature and Hubble 
constant, respectively. Eq. (3.7) shows that the smaller $t$ is, the
smaller $k$ will be. This is the so-called ``first out - last in" of
the evolution of density perturbations produced by the inflation \cite{KT}.

Using solution (2.20) and (3.7), the perturbations (3.6) can be
rewritten as
\begin{equation}
\left\langle \left ( \frac{\delta\rho}{\rho} \right )^2 \right \rangle_h
       \propto k^{-3\alpha/2},  \ \ \ \ {\rm if} \ \ \ k > k_e,
\end{equation}
where $k_e$ is the wavenumber of perturbations crossing out of horizon
at $t_e$. It is
\begin{equation}
k_e = 2\pi H\frac{T_0}{T_f} e^{H(t_e-t_f)}
\simeq 2\pi H\frac{T_0}{T_f} e^{-N}.
\end{equation}
Therefore, the primordial density perturbations produced during thermal
inflation are of power law with an index $-3\alpha/2$.  While in terms of
the density perturbations at a given time $t$, it can also be expressed as a
power law as \cite{LL}
\begin{equation}
\left\langle \left ( \frac{\delta\rho}{\rho} \right )^2 \right \rangle_t
       \propto k^{3+n},  \ \ \ \ {\rm if} \ \ \ k > k_e,
\end{equation}
and then the index $n$ becomes 
\begin{equation}
n= 1 -3\alpha/2.
\end{equation}
Consequently, the power spectrum of thermally induced perturbations is a
tilted power law.

The thermal scenario requires that all perturbations re-entering in the 
Hubble radius originate in the period of thermal inflation. 
The longest wavelength of the perturbation (3.8), {\it i.e,} $2\pi/k_{e}$,
should be larger than the present Hubble radius. In other words, 
$N$ should be large enough in order to have the inflated patch
cover the entire Hubble radius. From Eq. (3.9), we have then
\begin{equation}
N > \ln \left(\frac{HT_0}{H_0T_f}\right) \sim \ln \frac{T_0}{H_0} \sim 70.
\end{equation}
where we used $(T_0/H_0) \gg (T_f/H)$. Thus, from Eqs. (2.25) and (3.11), we
can find a lower limit to $n$ as
\begin{equation}
n_{\rm min} = 1 - 3\frac {\ln(T_f/H)}{\ln(T_0/H_0)}.
\end{equation}
Figure 3 plots $n_{\rm min}$ as a function of the inflation mass scale $M$ and 
$g_{\rm eff}$. $n_{\rm min}$ is sensitive to $M$, but
only slightly varying when $g_{\rm eff}$ is in the range $10^2 - 2\cdot10^3$.
Thus, as mentioned in Sec. II.B, the uncertainty given by $g_{\rm eff}$-factor
is not important.

We now try to find an upper limit to $n$. From Eqs. (2.9) and (2.25),
and $H^2 \simeq 8\pi V(\phi)/3 m_{\rm Pl}^2$, we have
\begin{equation}
N = \int_{\phi_e}^{\phi_f} \frac {H}{\dot{\phi}}d\phi=
  \frac{8\pi\gamma}{3m_{\rm Pl}^2} \int_{\phi_f}^{\phi_e} \frac{V}{V'}d\phi,
\end{equation}
where $\phi_e=\phi_i \exp (\alpha H t_e)$ and
$\phi_f=\phi_i \exp (\alpha H t_f) \sim \sigma / \sqrt{2}$. For the
potential (2.12), Eq. (3.14) gives
\begin{eqnarray}
N  & =  & \frac{2\pi \gamma}{3m^2_{\rm Pl}}
      \int_{\phi_f}^{\phi_e}
      (\phi - \frac{\sigma^2}{\phi})d\phi          \nonumber \\
       & = & 2\frac{\pi \gamma}{3m^2_{\rm Pl}}
    \left[\sigma^2 \alpha N - \frac {\sigma^2}{4} +
    \frac{\phi_e^2}{2}\right].
\end{eqnarray}
Because $\phi_e \ll \sigma/\sqrt{2}$, Eq. (3.15) can be rewritten as
\begin{equation}
N = \frac{\pi}{6} \frac{\sigma^2}{m_{\rm Pl}^2} \gamma(4\alpha N - 1)
\end{equation}
or
\begin{equation}
\alpha = \frac {1+ 6Nm_{\rm Pl}^2/\pi \sigma^2 \gamma}{4N} > 
\frac{3}{2\pi \gamma}\left( \frac{m_{\rm Pl}}{\sigma}\right)^2 > 
\frac{3}{2\pi \gamma},
\end{equation}
here we used $\sigma^2 = \lambda^{1/2}M^2 < m^2_{\rm Pl}$. Hence, an 
upper limit to $n$ is found to be
\begin{equation}
n < 1 - \frac{9}{4\pi\gamma}.
\end{equation}
 For most interesting cases, $\gamma \geq 70$ (see Sec. III.C), and
therefore, $n < 0.99$. Eq. (3.17) also provides another lower limit,
$\alpha > 1/4N$, and the corresponding upper limit to $n$ is
\begin{equation}
n \leq 1 - 3/8N.
\end{equation}
In the case of thermal inflation, $N \geq 70$ [Eq. (3.12)], and 
therefore,we obtain the same upper limit as Eq. (3.18), {\it i.e,} $n < 0.99$. 
Thus, roughly speaking, the allowed values of the power law index should fall
within the area between the curves of $n_{\rm min}$ and the line $n = 1$ in 
Fig. 3.

\subsection{Amplitudes of perturbations}

To calculate the amplitude of the perturbations we rewrite spectrum (3.8)
into
\begin{equation}
\left\langle \left ( \frac{\delta\rho}{\rho} \right )^2 \right \rangle_h
      = A \left( \frac{k}{k_0}\right )^{n-1},  \ \ \ \ {\rm if} \ \ \ k > k_e,
\end{equation}
where $k_0=2\pi H_0$. $A$ is the normalized amplitude of the perturbation on 
scale $k=k_0$, which re-enters the Hubble radius $1/H_0$ at present time.
 From Eqs. (3.6), and (3.7), we have
\begin{equation}
A = \frac{405}{(2\pi)^3}\frac{\Gamma_{\phi}}{g_{\rm eff}H}
     \left(\frac{H_0T_f}{HT_0}\right)^{n-1}\left(\frac{H}{T}\right)^{3}
    e^{(n-1)H(t_f - t)}.
\end{equation}
 Applying Eq. (2.20), the radiation temperature at the horizon-crossing 
moment, $t$, can be expressed as $T(t) = T_f \exp[(n-1)H(t_f - t)/3]$.  With 
the help of Eq. (2.25), we obtain 
\begin{eqnarray}
A & = & \frac{405}{(2\pi)^3}\frac{\Gamma_{\phi}}{g_{\rm eff}H}
     \left(\frac{H_0}{T_0}\right)^{n-1}\left(\frac{T_f}{H}\right)^{n-4}
       \nonumber \\
   & = &  \frac{405}{(2\pi)^3}\frac{\Gamma_{\phi}}{g_{\rm eff}H}
     \left(\frac{H_0}{T_0}\right)^{n-1}
     e^{-(1-n)(4-n)N/3}.
\end{eqnarray}
On the other hand, Eqs. (2.23), (2.24), and (2.25) render
\begin{equation}
\frac{\Gamma_{\phi}}{H} =
  \frac {9a(\frac{H}{M})^4 e^{4(1-n)N/3}}
  {(1-n) - 3a(\frac{H}{M})^4 e^{4(1-n)N/3}}.
\end{equation}
Substituting Eq. (3.23) into Eq. (3.22), we have finally
\begin{equation}
A =  \frac{405}{(2\pi)^3}\frac{9\pi^2}{30}
     \left(\frac{H_0}{T_0}\right)^{n-1}
     \frac {(\frac{H}{M})^4 e^{(1-n)nN/3}}
     {(1-n) - 3a(\frac{H}{M})^4 e^{4(1-n)N/3}}.
\end{equation}
Because for a given energy scale of inflation $M$, power law index $n$
and $g_{\rm eff}$, $N$ can be approximately determined by Eq. (2.25), the
amplitude $A$ is a function of $M$, $n$ and $g_{\rm eff}$.  Figure 4 shows 
$A$ as a function of $n$ for given parameters $M$ and $g_{\rm eff}$.  The 
values of $n$ in Fig. 4 cover the entire allowed range shown in Fig. 3.  Again,
it can be seen from Fig. 4 that $g_{\rm eff}$ gives only a slight effect on the
magnitude of $A$.  But the magnitude of $A$ will change by about 8 orders when
$n$ changes from 1 to 0.60.

To check the consistency of these solutions, we calculate $\Gamma_{\phi}$
as function of $M$, $n$ and $g_{\rm eff}$. The result is plotted in Fig. 5.
Figure 5 shows that in all cases $\Gamma_{\phi}$ will be in the range 
around $10^2H$.  These values are satisfied very well the condition of
$H<\Gamma_{\phi}<M$.  Figure 5 also shows that in the parameter range allowed 
by the thermal scenario the values of $\Gamma_{\phi}$ are mainly determined 
by $M$, but not $n$ and $g_{\rm eff}$.

To replace $n$, $A$ and $\Gamma_{\phi}$ can also be treated as functions of
$M$, $N$, and $g_{\rm eff}$.  Figures 6 and 7 give the relationships of $A$ - 
$N$ and $\Gamma_{\phi}$ - $N$, respectively. Similarly, the variation of
$g_{\rm eff}$ does not have much effect on the amplitudes of primordial density
perturbations, and the friction $\Gamma_{\phi}$. 

\subsection{Suppression on super-Hubble scales}

As in the standard scenario, the fluctuations of $\phi$ field in the thermal 
scenario leads to curvature perturbations.  However, the radiation $\rho_r$ in 
the thermal scenario will also contribute to the total perturbations.  Instead
of the adiabatic mode, the fluctuation in $\rho_r$ will provide
an isocurvature component of perturbations.  This is because the relaxation 
time-scale of the radiation during the thermal inflation is equal to or less 
than $1/\Gamma_{\phi}$, and much shorter than $1/H$.  Namely, the radiation 
component is always in thermal equilibrium, and their thermal (temperature) 
fluctuation is much less than the fluctuations caused by $\phi$ field phase 
transition (fluctuation-dissipation relation).  Therefore, in terms of the 
equation of state, the compound system of $\phi$ field and radiation is 
non-uniform.  This is isocurvature initial condition which specifies the 
relative abundance of the $\phi$-decayed particles and the thermal component. 

One can simply estimate the amplitude of the isocurvature perturbations 
by the following way \cite{PAD}.  As has been discussed in Sec. II, the 
inflation ended  at the
stage of $\rho_r \sim \rho_{\phi}$.  Hence, at $t_f$ when the radiation 
dominated phase starts, the total density contrast should be
\begin{equation}
\delta = \frac{\delta \rho}{\rho}= 
\frac {\delta \rho_{\phi} + \delta \rho_r}{\rho_{\phi} + \rho_r}
 \simeq \frac{1}{2} \frac{\delta \rho_{\phi}}{\rho_{r}}
\end{equation}
The amplitude of the entropy perturbations is then given by 
\begin{equation}
S = \frac{3}{4}\frac{\delta \rho_{\phi}}{\rho_{\phi}} - \frac{3}{4} \delta_r
  \simeq \frac{3}{8} \frac{\delta \rho_{\phi}}{\rho_{r}}
  \simeq \frac{3}{4} \frac{\delta \rho}{\rho}
\end{equation}
These isocurvature fluctuations will eventually lead to density perturbations
of amplitudes $S$ once they enter the horizon.

Accordingly, the amplitude of density perturbations before re-entering the
horizon should be lower than the re-entered perturbations by a factor of about
$1 + 3/4 = 1.75$.  In other words, the density perturbations are suppressed
on super-Hubble scales. The power spectrum should therefore be modified to
\begin{equation}
P(k) \equiv (\delta\rho/\rho)^2_{t}=A \left(\frac{k}{k_0}\right)^{3+n}f(k) .
\end{equation}
where the function $f(k)$ is
\begin{equation}
f(k) = \left\{ \begin{array}{ll}
               1                 & \hspace{1cm} k > k_0 \\
               1 - s             & \hspace{1cm} k \leq k_0
              \end{array}
      \right.
\end{equation}
where $s = 1 - (1.75)^{-1} \sim 0.4$ is called the super-Hubble suppression 
factor.

\section{Discussions and conclusions}

Assuming that the initial inhomogeneity of the universe originated 
in the period of thermal inflation, the properties of the mass density 
perturbations have been studied. A number of the features of the 
perturbations have been accounted for, including (1) the spectrum of the 
perturbations is of power law, but titled, and the allowed values of 
the power law index are in the range of about $0.6 < n < 1$; (2) for a given 
mass scale of the inflation, the amplitude of the perturbations $A$ is
almost completely determined by the index $n$; (3) the perturbations
are suppressed on super-Hubble scales. 

In this analysis the popular $\phi^4$ potential of the inflationary
field was employed. However, only one parameter, the mass
scale of the inflation $M$, is involved in the predictions listed above,
and no other details of the model are needed. Therefore, we believe that 
these qualitative features should be common for the thermal
inflation scenario. Actually, the thermal model is based on two 
thermodynamical requirements: (a) a thermalized heat bath exists
during inflation and (b) the initial fluctuations are given by the
fluctuation-dissipation theorem. Like many thermodynamical problems,
the results are mainly determined by the thermodynamical conditions,
and less dependent on other details. 

The predictions (1) and (2) can be tested by current observations of the
temperature fluctuations of the Comics Microwave Background (CMB).
The four-year COBE-DMR observations found $n \sim 1.2 \pm 0.3$,
$Q_{rms-PS} \sim 15.3^{+3.7}_{-2.8} \mu K$, or $Q_{rms-PS} \sim 18 
\mu K$ for fixed $n=1$ \cite{cobe}. We have plotted this observational
allowed area of ($A$, $n$) in Fig. 4. Meanwhile, the relation between 
$n$ and $Q_{rms-PS}$, i.e. $\delta Q/Q = -(\ln 5)\delta n$ from data fitting,
has also been plotted in Fig. 4. All these results are consistent with 
the thermal scenario if the mass scale of the inflation is equal to
 about $10^{15}$ GeV.

However, one should be careful in the comparison of the thermal inflation 
predictions with the COBE results. In all the fitting of the COBE-DMR 
observations done so far, only the spectra described by two parameters 
($n$, $Q_{rms-PS}$) or ($n$, $A$) has been considered. In other words, 
the published results of the COBE data analysis provide only the most 
likely area in the bi-parameter ($n$, $Q_{rms-PS}$) space. This fitting
implies that it was assumed that there is no suppression on scales larger than
the Hubble radius, or $s=0$ in Eqs. (3.27) and (3.28).

To test the predictions of thermal inflation, we need to fit the data 
with spectrum suppressed on the super-Hubble scales. The most likely
values of $n$ and $Q_{rms-PS}$ (or $n$ and $A$) are generally dependent
on the suppression on super-Hubble scales. Because long wave contributions
are reduced by the super-Hubble suppression, one can expect that the
most likely value of $n$ given by fitting with the suppression will 
be smaller than that without the suppression \cite{JF}. The result would
then be moving toward the direction in favor of a tilted spectrum.
Nevertheless, the suppression should not dramatically change the allowed
area shown in Fig. 4. Therefore, one may already conclude that for
thermal inflation, the most likely value of the inflationary mass scale
is about 10$^{15}$ GeV, and the regions larger than $10^{16}$ GeV or
lower than $10^{14}$ GeV are likely to be excluded.

For a better test, we need a tri-parameter fit to find the most likely values 
of $n$, $A$, and $s$. Various models of the perturbation origins are difficult 
to distinguish in the bi-parameter $n$ and $Q_{rms-PS}$ ($A$) space.
All models of the ``standard" inflation, thermal inflation, and phase 
transition defects, can produce the perturbation spectrum with about 
same bi-parameters. However, the super-Hubble suppression factor $s$ is
different for different models. In the ``standard" inflation model,
the perturbations are assumed to be adiabatic, and then $s=0$. On the 
other hand, isocurvature perturbations require $s=1$. For models
such as defects in phase transitions, causality allows no perturbations
on scales larger than the Hubble radius, we have $s=1$ too. The thermally
originated fluctuations are neither purely adiabatic, nor purely 
isocurvature. Because the thermal inflation ends at the time when
the energy of the radiation is comparable to the $\phi$ field energy,
the suppression factor $s$ should not be restricted to 0 or 1. Therefore, 
the value of $s$ is crucial to distinguish the thermal model from
the others.

\acknowledgments
Wolung Lee would like to thank Jesus Pando, Michael Shupe, Hung Jung Lu, and 
Charles Liu for helpful discussions.

\newpage

\begin{figure}

\caption{Numerical solutions of the evolutions of $\phi$ field and
radiations in a thermal inflation model with $V(\phi)=\lambda (\phi^2 -
\sigma^2)^2$ at the inflation mass scale $M = 10^{15}$ GeV. The parameters are 
taken to be $\lambda=4\cdot 10^{-17}$, $\Gamma_{\phi}=15H_i$, and 
$g_{\rm eff}=100$. The dot-dashed and solid lines
are the evolutions of $H(t)$ and the temperature $T$ of the radiation,
respectively. The circles are given by the approximate solution of $T$,
which perfectly matches the exact solution by the end of the slow-roll era.
$t_b$ is the time at which the temperature rebounds, and $t_f$ is the end of
inflation. $T$ and $H$ are in units of $H_i\equiv [8\pi V(0)/3m_{\rm Pl}^2]^
{1/2}$, and $t$ is in units of $1/H_i$.}
\label{1}
 
\end{figure}

\begin{figure}

\caption{An exact solution of $H$ and $T$ similar to that shown
in Figure 1, but the parameters are taken to be $\lambda=4\cdot 10^{-17}$,
$\Gamma_{\phi} =150H_i$, $g_{\rm eff}=100$. $t_e$ is the time of the onset
of thermal inflation.}
\label{2} 
\end{figure}

\begin{figure}

\caption{The lower limit of the spectral index $n$ as a function of the 
mass scale $M$. $g_{\rm eff}$ is taken to be 100 (dotted line), 500 
(dot-dashed line) and 2000 (solid line), respectively.}
\label{3}

\end{figure}

\begin{figure}

\caption{The amplitudes of the power spectrum as a function of $n$ which 
covers the allowed range $n_{\rm min} < n <1$. The
mass scales $M$ are taken to be $10^{14}$, $10^{15}$ and $10^{16}$ GeV
labeled at the three sets of curves, and $g_{\rm eff}$ are 100 (dotted lines),
500 (dot-dashed lines) and 2000 (solid lines). The shaded zone is the
allowed area of the values ($n$ and $A$) given by the COBE-DMR data.
The solid line within the shaded zone represents the approximated relation
$\delta Q/Q = -(\ln 5)\delta n$.}
\label{4}

\end{figure}

\begin{figure}

\caption{$\Gamma_{\phi}$ as a function of spectral index $n$, which 
covers the allowed range $n_{\rm min} < n <1$.  The
mass scales $M$ are taken to be $10^{14}$, $10^{15}$ and $10^{16}$ GeV
labeled at the three sets of curves, and $g_{\rm eff}$ are 100 (dotted lines),
500 (dot-dashed lines) and 2000 (solid lines). $\Gamma_{\phi}$ is in 
units of $H_i$.}
\label{5}

\end{figure}

\begin{figure}

\caption{The amplitude of the power spectrum as a function of the 
thermal duration $N$ in the 
range of $N > \ln(T_0/H_0)\sim 70$, {\it i.e,} the range of thermal inflation.
The mass scales $M$ are taken to be $10^{14}$, $10^{15}$ and $10^{16}$ GeV
labeled at the three sets of curves, and $g_{\rm eff}$ are 100 (dotted lines),
500 (dot-dashed lines) and 2000 (solid lines).}
\label{6}
\end{figure}

\begin{figure}

\caption{$\Gamma_{\phi}$ as a function of the thermal duration $N$ in the 
range of $N > \ln(T_0/H_0) \sim 70 $. The mass scales $M$ are taken to be
$10^{14}$, $10^{15}$ and $10^{16}$ GeV labeled at the three sets of curves,
and $g_{\rm eff}$ are 100 (dotted lines), 500 (dot-dashed lines) and 2000
(solid lines). $\Gamma_{\phi}$ is in units of $H_i$.}
\label{7}

\end{figure}

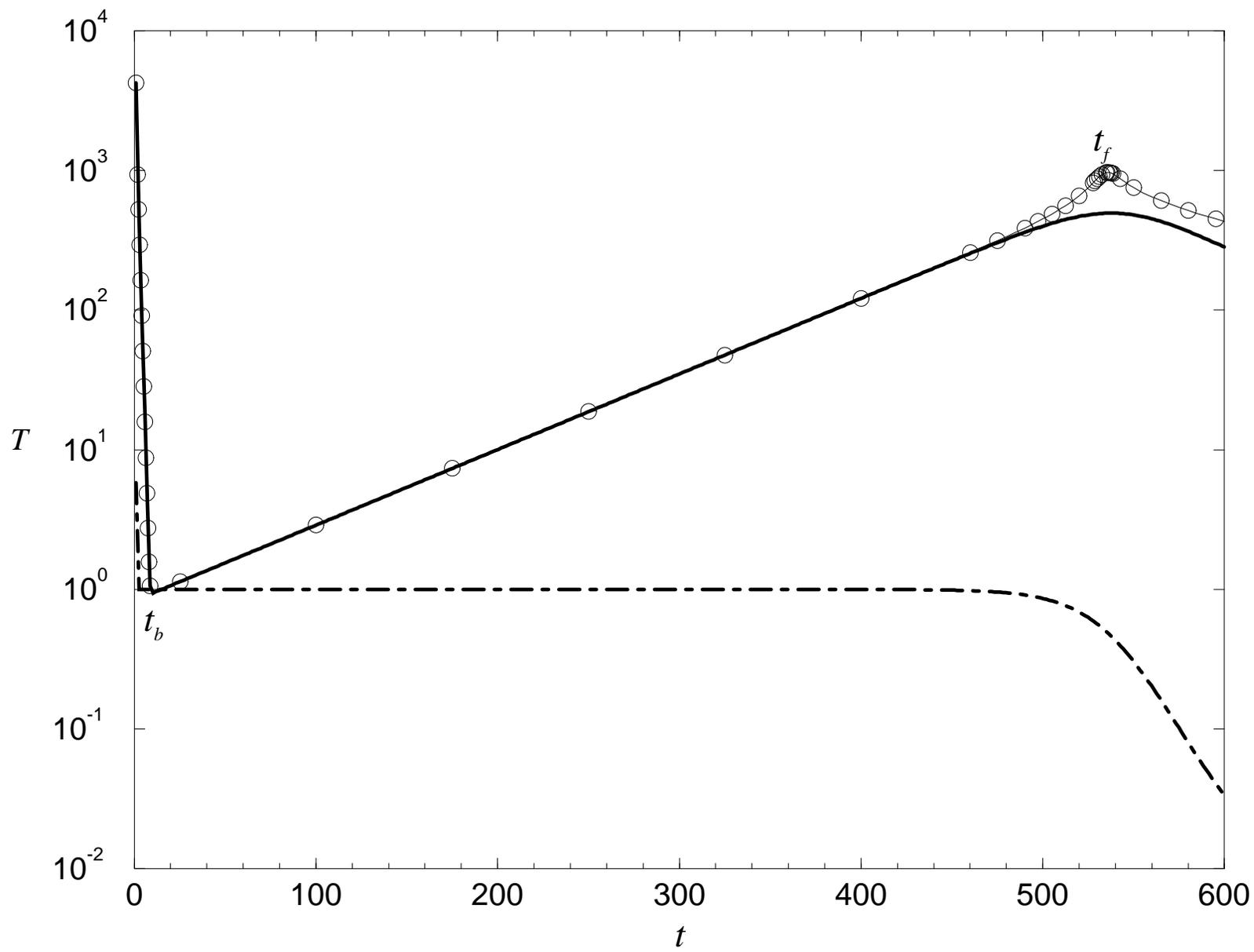
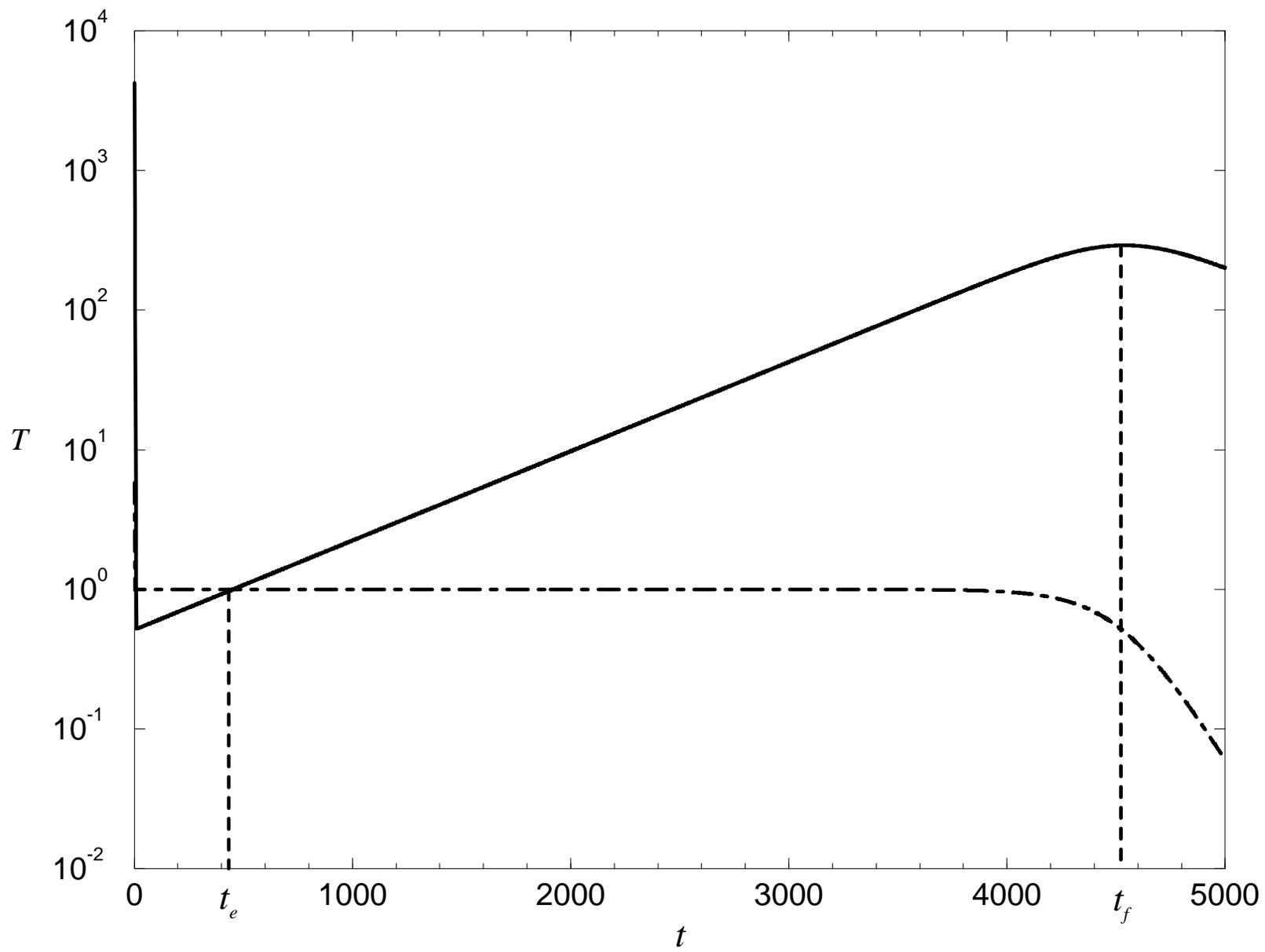
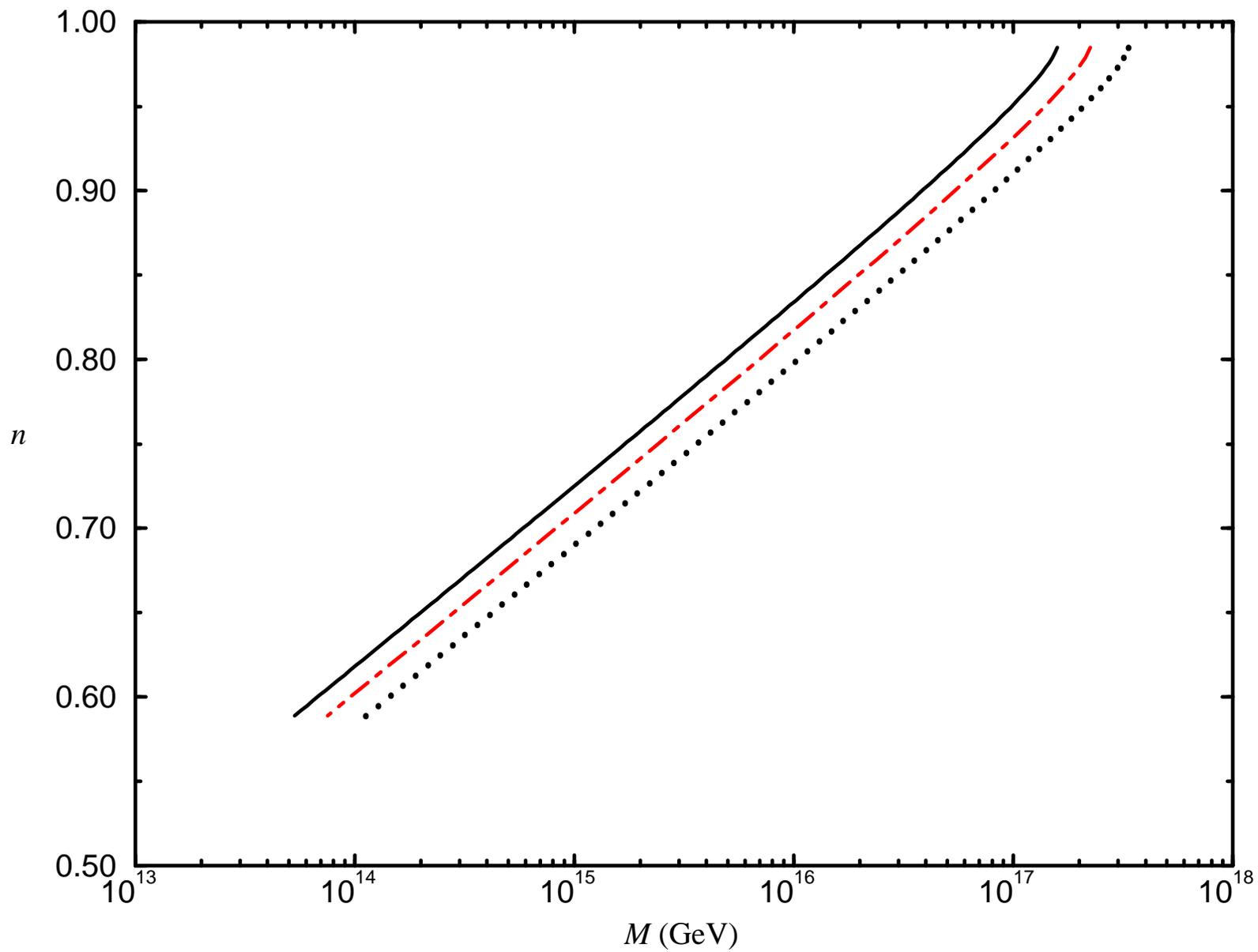
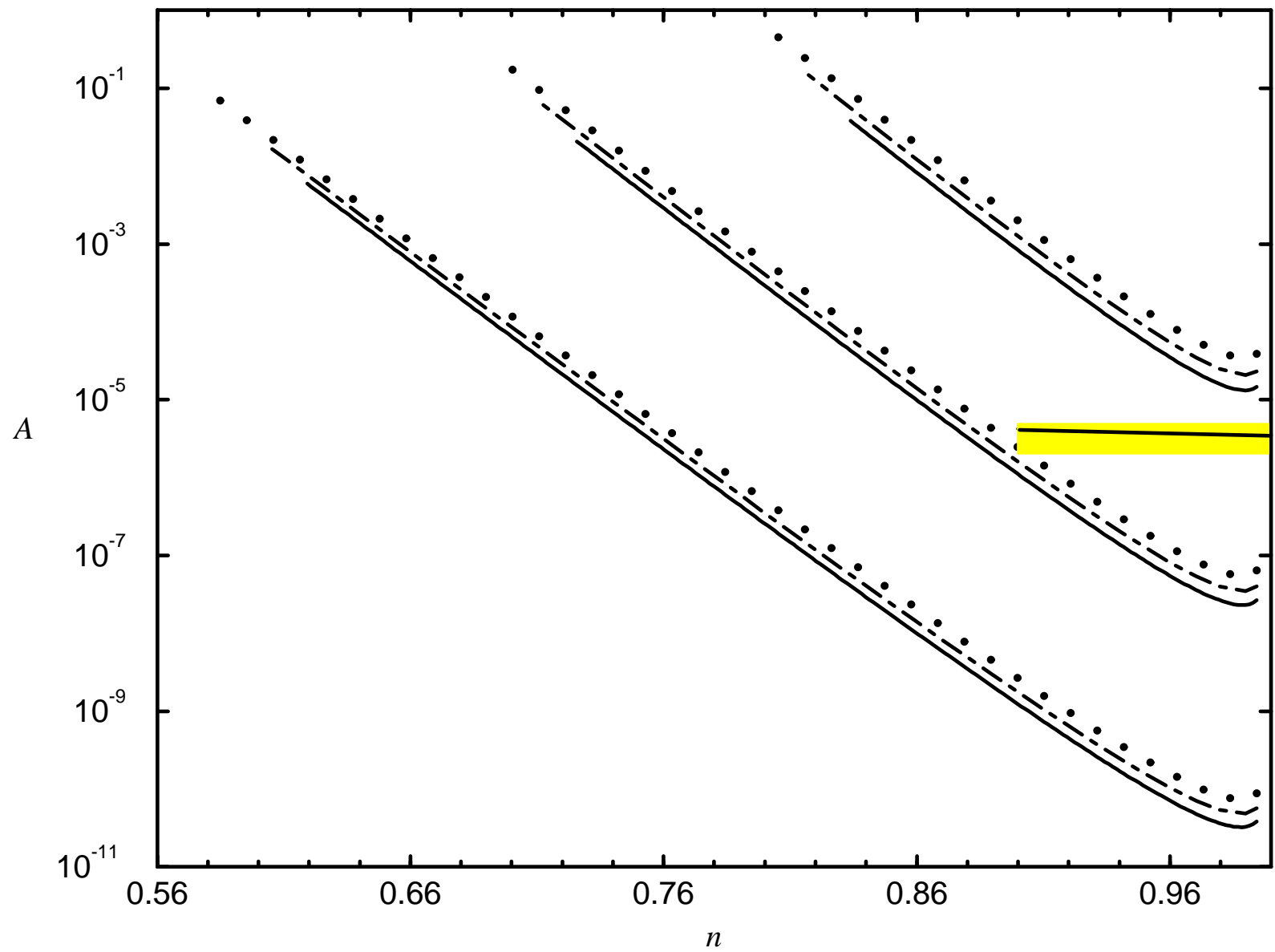
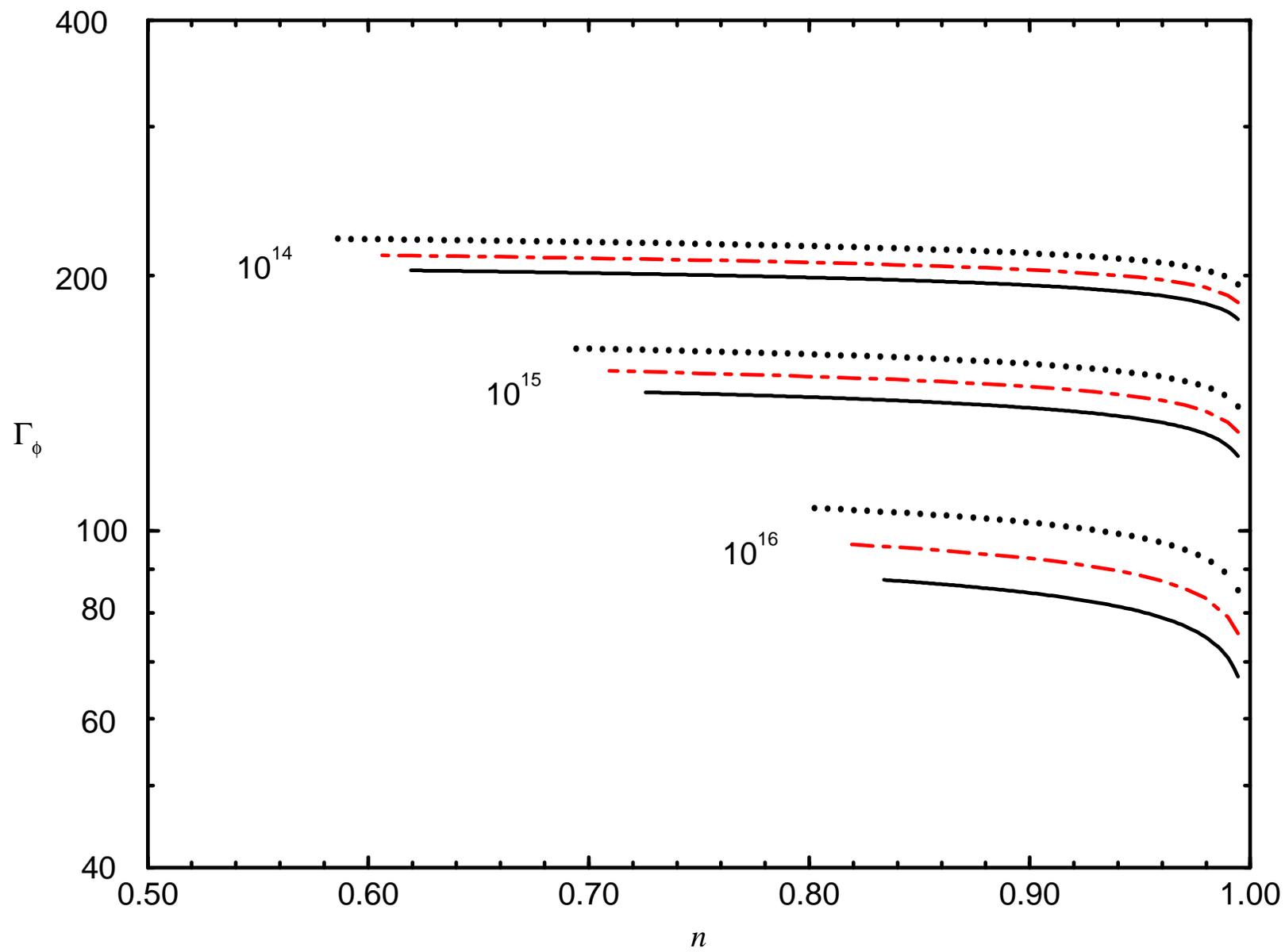
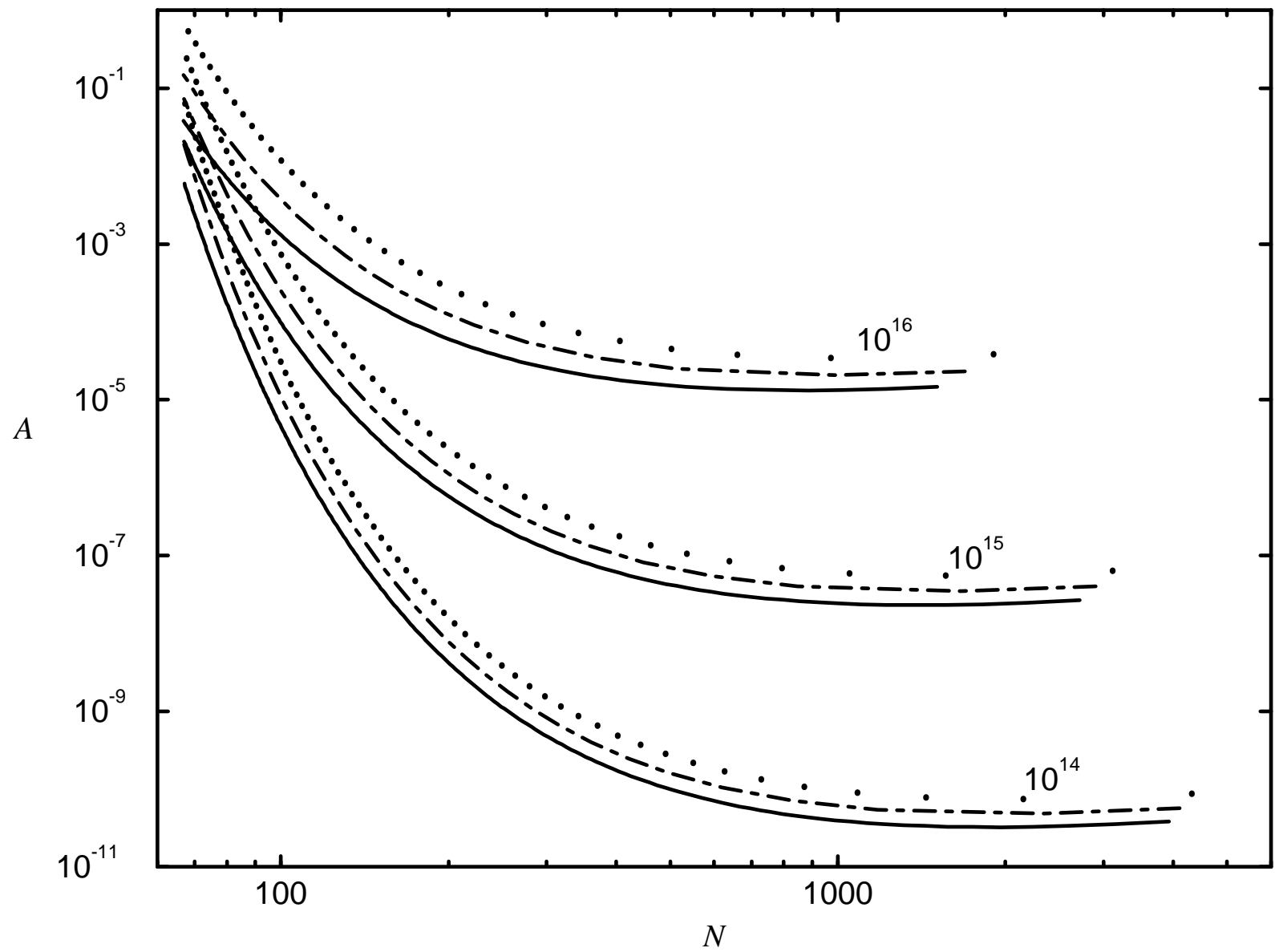
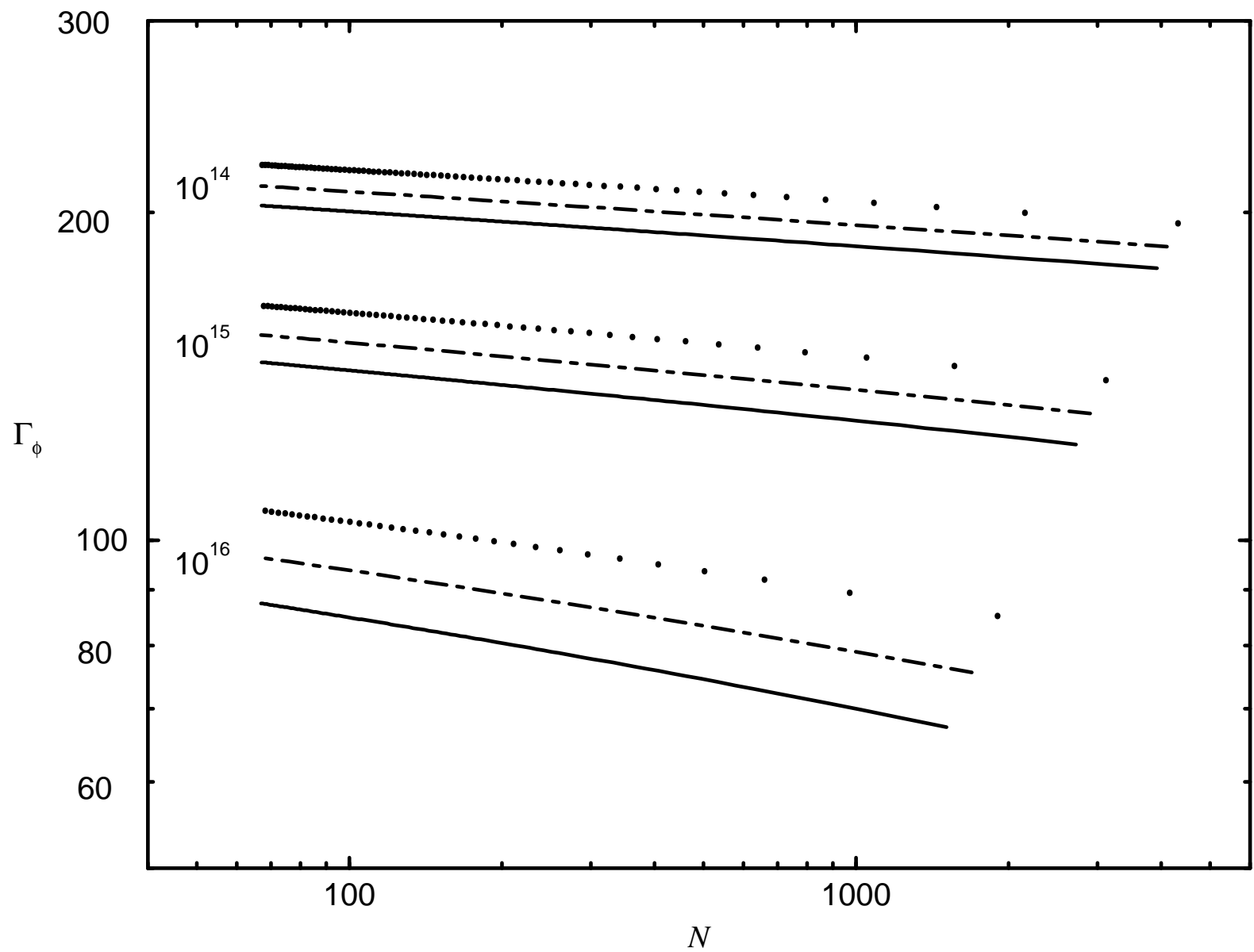
\end{document}